\documentclass[twocolumn,english,aps,prl,reprint,nofootinbib,superscriptaddress]{revtex4-2}

\pdfoutput=1
\usepackage[T1]{fontenc}
\usepackage[utf8]{inputenc}
\usepackage{amsmath}
\usepackage{amssymb}
\usepackage{graphicx}
\usepackage[caption=false]{subfig}
\usepackage{color}
\usepackage{float}
\usepackage{tabularx}
\newcolumntype{Y}{>{\centering\arraybackslash}X}
\usepackage{multirow}
\usepackage{hyperref}
\hypersetup{
    colorlinks,
    linkcolor={black},
    citecolor={black},
    urlcolor={blue},
}
\makeatletter

\@ifundefined{textcolor}{}
{%
 \definecolor{BLACK}{gray}{0}
 \definecolor{WHITE}{gray}{1}
 \definecolor{RED}{rgb}{1,0,0}
 \definecolor{GREEN}{rgb}{0,1,0}
 \definecolor{BLUE}{rgb}{0,0,1}
 \definecolor{CYAN}{cmyk}{1,0,0,0}
 \definecolor{MAGENTA}{cmyk}{0,1,0,0}
 \definecolor{YELLOW}{cmyk}{0,0,1,0}
 \definecolor{PURPLE}{rgb}{0.7,0,0.7}
 \definecolor{dgreen}{rgb}{0,0.6,0}
}

\usepackage{pslatex}
\usepackage{babel}

\makeatother

\begin{document}

\title{Thermal Emission of Strontium in a Cryogenic Buffer Gas Beam Source}

\author{Andrew Winnicki}
\email{andrew.winnicki.sc@gmail.com}
\affiliation{Harvard-MIT Center for Ultracold Atoms, Cambridge, MA 02138, USA}
\affiliation{Department of Physics, Harvard University, Cambridge, MA 02138, USA}

\author{Zack D. Lasner}
\affiliation{Harvard-MIT Center for Ultracold Atoms, Cambridge, MA 02138, USA}
\affiliation{Department of Physics, Harvard University, Cambridge, MA 02138, USA}

\author{John M. Doyle}
\affiliation{Harvard-MIT Center for Ultracold Atoms, Cambridge, MA 02138, USA}
\affiliation{Department of Physics, Harvard University, Cambridge, MA 02138, USA}

\date{\today}

\begin{abstract}

We demonstrate production of cold atomic strontium (Sr) and strontium-containing molecules (SrOH) in a cryogenic buffer gas beam source via direct heating of strontium oxide (SrO) with 30~mJ laser pulses several milliseconds long. $3.7(2)\times10^{14}$ Sr atoms are released, which represents a factor of 7 increase in atomic production per pulse compared to nanosecond-scale ablation laser pulses. A peak atomic density of $1.93(6) \times 10^{12}$ atoms/cm$^3$ is achieved, which corresponds to a factor of 2 increase relative to ablation. We further propose extensions of this method to other atomic and molecular species.

\end{abstract}

\maketitle

\section{Introduction}

Cryogenic buffer gas beam (CBGB) sources provide slow, high-flux, cold beams of atoms and molecules~\cite{hutzler_buffer_2012}. CBGBs are used in a wide range of scientific applications including molecular laser cooling and searches for CP violating physics beyond the Standard Model \cite{augenbraun_laser-cooled_2020,acme_collaboration_improved_2018}. CBGBs produce both atoms and molecules, generated either directly through ablation or through cold chemical reactions between atoms and reactant molecules in the cryogenic buffer gas cell. While certain atomic and molecular species can be introduced into a cryogenic source via a capillary heated to room temperature or below, nearly all species of interest have a low vapor pressure below temperatures of $\sim$100$^{\circ}$~C, for which heat loads from a capillary can be prohibitively challenging to manage~\cite{patterson_intense_2009, piskorski_cooling_2014, porterfield_high_2019}. Pulsed lasers with durations of $\sim$10~ns and $\sim$30~mJ of energy are therefore commonly employed to produce a wide variety of atoms and molecules in CBGBs, irrespective of the precursor material vapor pressure. For favorable precursors, production yields as large as $\sim$$10^{13}$ atoms/pulse or $\sim$$10^{10}$ molecules/pulse are typical~\cite{hutzler_buffer_2012}.

Most applications would benefit from still higher production yields. For example, the sensitivity of a Ramsey-type precision measurement increases with the square root of the number of probed atoms or molecules, with the potential for further improvements using established entanglement methods ~\cite{acme_collaboration_methods_2017,bilitewski_dynamical_2021,miller_two-axis_2024}. For quantum simulation and quantum computing applications, increased atomic and molecular production yields can benefit experiments by facilitating the robust produciton of atomic or molecular arrays~\cite{acme_collaboration_methods_2017, endres_atom-by-atom_2016,anderegg_optical_2019,kaufman_quantum_2021} and increasing the phase-space density of a trapped sample~\cite{cheuk_lambda-enhanced_2018, wu_high_2021, hallas_high_2024}. One way to increase the number of useful molecules in an experiment is at the very beginning of the experimental sequence---that is, by increasing the atomic and molecular production yield. Thus, more efficient atomic and molecular production methods are of great interest.

Earlier experiments have demonstrated that strontium can be emitted from the surface of a bulk SrO target by laser heating to load a magneto-optical trap (MOT)~\cite{kock_laser_2016, he_towards_2017}. Later experiments showed that this laser heating method could also be applied to granules of elemental strontium to load a MOT~\cite{hsu_laser-induced_2022}. Here, we extend this approach to a CBGB source, demonstrating favorable atomic yield compared to ablation. We also establish evidence supporting a thermal emission mechanism, in contrast to well-established laser-induced atomic desorption phenomena in alkali metals~\cite{meucci_light-induced_1994, gozzini_light-induced_1993, rebilas_comment_2010, rebilas_light-induced_2009, barker_light-induced_2018, alexandrov_light-induced_2002, anderson_loading_2001, zhang_light-induced_2009}.

Additionally, we report production of SrOH molecules by introducing water vapor into the cell during atomic emission. Strontium-containing molecules, most notably SrF and SrOH, are of particular interest due to their suitability for laser cooling and their applications to both quantum information science~\cite{carr_cold_2009, langin_toward_2023, micheli_toolbox_2006, demille_quantum_2002, yu_scalable_2019} and precision measurement~\cite{kozyryev_enhanced_2021, kozyryev_precision_2017, frenett_vibrational_2024}. Both molecules can be produced in CBGBs via ablation of Sr metal in the presence of a reagent gas (either SF$_{6}$ or H$_{2}$O)~\cite{lasner_vibronic_2022, jorapur_high_2024}. Increasing the total atomic yield, or achieving the same yield with lower heat load on the cryogenic system, may therefore prove favorable for molecular production. Finally, we propose methods of producing other chemical species via direct laser heating of precursor materials.

\section{Methodology}

\begin{figure}
    \centering
    \includegraphics[width=\columnwidth]{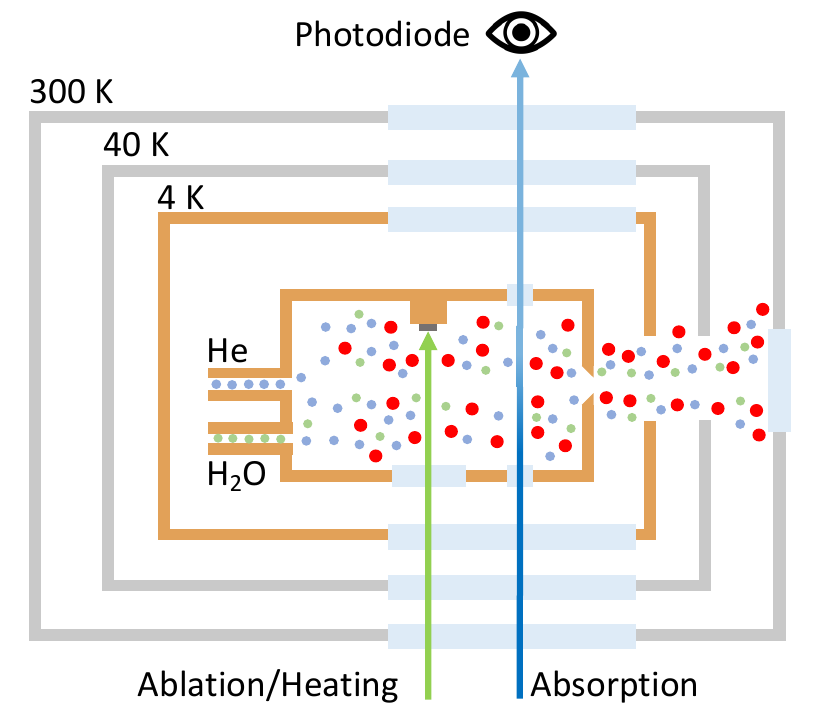}
    \caption{Experimental setup. Sr atoms are produced via ablation or laser heating of a strontium oxide or metal target, and cooled via collisions with cold helium buffer gas. Water is optionally introduced to the cell via a separate heated capillary to produce SrOH molecules. Laser absorption near the exit of the cell measures production yields.}
    \label{fig:experimental_setup_schematic}
\end{figure}

Sr atoms and SrOH molecules are produced in a CBGB source~\cite{hutzler_buffer_2012}; see Figure~\ref{fig:experimental_setup_schematic}. The CBGB apparatus consists of a copper cell, cooled to 4~K, with a 43~mm inner diameter bore and 9~mm output aperture. Optical access for laser heating or ablation and for absorption measurements is provided by float glass windows. Strontium atoms are released from a target, described below, via either a high-power "heating" laser or a pulsed Nd:YAG  "ablation" laser. The production of SrOH molecules from gaseous Sr atoms is induced by flowing water vapor through a capillary heated to 10$^{\circ}$~C. The water molecules mix and interact with the strontium atoms, undergoing a chemical reaction to produce SrOH. Helium is introduced at a rate of 9.85~sccm to facilitate thermalization of the gaseous Sr and SrOH molecules with the cell, and to extract those species through the cell aperture.

The heating laser consists of a 20~W, 532~nm diode-pumped solid state laser. The output of the heating laser is directed through a mechanical shutter to coarsely control the pulse length and minimize thermal lensing in subsequent optics. The laser then passes through a 0.5$\times$ telescope and then a quartz acousto-optic modulator (AOM) for precise pulse duration control. The first-order diffracted beam is then directed through a 10$\times$ telescope and a 400~mm achromatic doublet lens, which focuses the beam to a measured waist of 50~$\mu$m. The laser is directed onto a pressed powder target consisting of a 2:3 mixture (by volume) of HfC and SrO, respectively. Both powders are 325 mesh size (44~$\mu$m diameter granules or smaller). HfC is a highly thermally insulating and refractory ceramic, which serves to darken the pressed powder target and significantly improve the absorption of incident laser power. Aside from facilitating absorption of laser light, we do not expect the HfC particles to play any fundamental role in the production of gas-phase Sr atoms.

For establishing an ablation reference, an Nd:YAG laser, frequency-doubled to 532~nm, is focused onto a Sr metal sample mounted directly adjacent to the pressed powder target. Ablating strontium metal results in a substantially higher flux of atoms than what is achieved by ablating the pressed powder targets. The total ablation pulse energy is set to 30 mJ, a typical value for molecular beam experiments.

To measure the production of atoms in the cell, an external cavity diode laser (ECDL) operating at the 461~nm $^1S_0 - ^1P_1$ atom transition is directed through the absorption windows near the exit of the cell. The atomic density is inferred from the attenuation of the laser light. The molecular production is likewise measured via absorption of the 688~nm $\tilde{X}^{2}\Sigma - \tilde{A}^{2}\Pi_{1/2}$ photon cycling transition of SrOH, generated by an ECDL. For all atomic data, we convert the measured absorption laser optical depth to atom density using the Beer-Lambert law for the Sr $^1S_0 - ^1P_1$ transition. The absorption laser is blue-detuned from the resonance for $^{88}$Sr by typically 2~GHz to avoid saturating the laser attenuation. We account for the detuning from resonance for each isotope, weighted by natural abundances, when computing the atomic density. We measure molecular absorption on resonance with the transition for $^{88}$SrOH.

To better understand the atomic emission process, we probe atom production at a variety of total deposited pulse energies and heating laser powers. For a given total energy and laser power, the pulse duration is set appropriately by switching the RF input to the AOM. We probe molecular production with 30~mJ deposited pulse energy under the conditions found to maximize the peak atomic density.

\section{Results}

\begin{figure}
    \centering
    \includegraphics[width=\columnwidth]{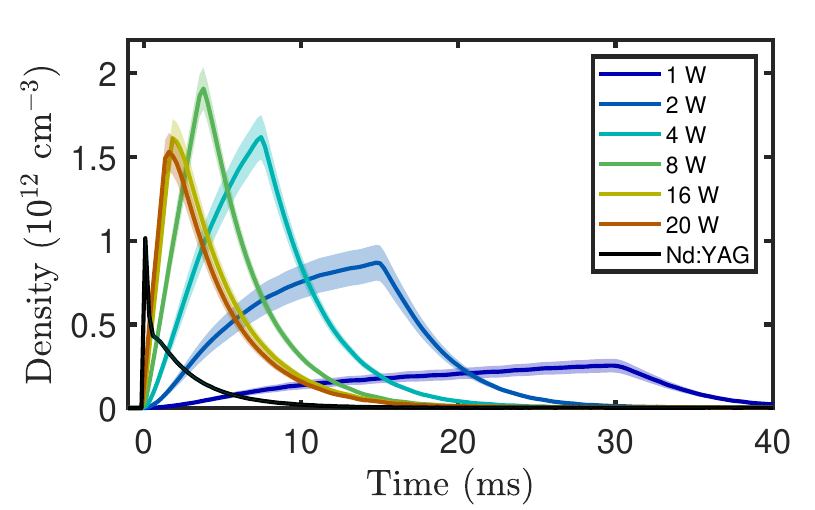}
    \caption{Time traces of atomic emission. Laser pulse energy is fixed at 30 mJ. The black trace shows Nd:YAG ablation yield, while other traces show laser heating at various powers. Filled area shows $\pm2\sigma_{\bar{\rho}}$, where $\sigma_{\bar{\rho}}$ is the standard error of the mean density at a given time.}
    \label{fig:time_traces}
\end{figure}

To compare thermal emission and ablation, we employ 30~mJ laser pulses, similar to typical ablation-based  sources. In Figure~\ref{fig:time_traces}, we show time traces of the measured atomic density. As the heating laser power is increased, the slope of atomic density vs. time increases correspondingly, but the constraint of a fixed pulse energy necessitates a shorter pulse duration at higher power. Also shown is a reference 30~mJ laser pulse ablation trace. The ablation laser was directed onto a Sr metal target.

\begin{figure}
    \centering
    \includegraphics[width=\columnwidth]{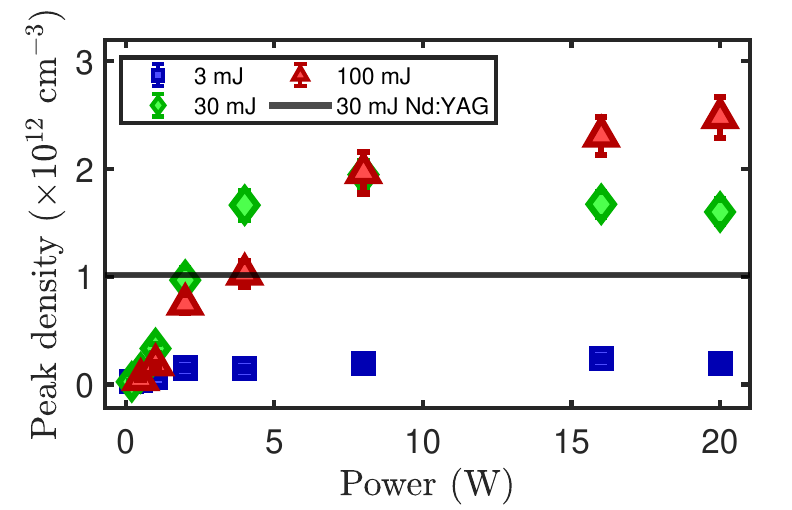}
    \caption{Peak in-cell atomic density for 3~mJ, 30~mJ, and 100~mJ laser pulses, as a function of heating laser power. Error bars show 95\% confidence intervals. Horizontal black line is the peak density of the 30~mJ Nd:YAG pulse on Sr metal.}
    \label{fig:peak_density}
\end{figure}

In Figure~\ref{fig:peak_density}, we show the peak atomic density as a function of laser power with 3~mJ, 30~mJ, and 100~mJ laser pulses. At 30 mJ pulse energy and 8~W of laser power (3.75~ms pulse duration), the peak density reaches a maximum of $1.93(6) \times 10^{12}$ atoms/cm$^3$, approximately $2\times$ that achieved via ablation ($1.01(1) \times 10^{12}$ atoms/cm$^3$). At either higher or lower laser powers, a lower peak density is achieved. The existence of an optimal laser power can be understood from the competition between two effects. At higher laser powers, the atomic density tends to increase more quickly with respect to heating time. On the other hand, the constraint of a fixed total deposited energy requires that the pulse duration is shorter at higher laser power.

\begin{figure}
    \centering
    \includegraphics[width=\columnwidth]{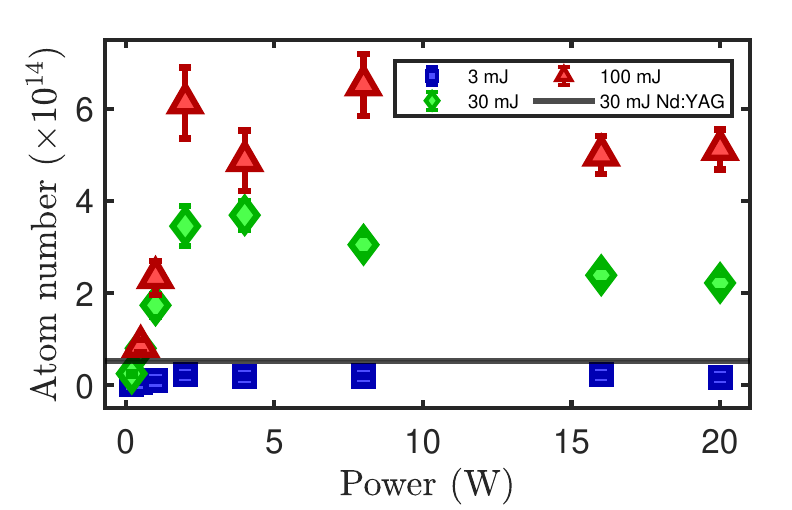}
    \caption{Number of atoms produced for 3~mJ, 30~mJ, and 100~mJ laser pulses, as a function of heating laser power. Error bars show 95\% confidence intervals. Horizontal black line denotes the number of atoms produced by the 30~mJ Nd:YAG pulse on Sr metal.}
    \label{fig:integrated_signal}
\end{figure}

In Figure~\ref{fig:integrated_signal}, we show the total number of atoms produced as a function of laser power for pulses with 3~mJ, 30~mJ, and 100~mJ of total energy. We also show the ablation production optimized at 30~mJ of pulse energy for reference. The atom number is obtained by integrating the atomic density over time (see App.~\ref{sec:atom_number} for details). For 30~mJ of energy, atomic production is maximized around 2$-$4~W of heating power, reaching a value of $3.7(2)\times10^{14}$ atoms, which is 7$\times$ more than the 30 mJ ablation reference ($0.54(1)\times10^{14}$ atoms). The gain in atom number is larger than the gain in peak density due to the non-negligible pulse duration ($\sim$10~ms), whereas the ablation peak is extremely narrow in time. As expected, the atom number grows with deposited energy, but the peak value for 100~mJ of energy is less than twice that for 30~mJ of energy.

We observed SrOH molecule production with thermal emission by flowing water vapor into the cryogenic cell through a heated capillary~\cite{lasner_vibronic_2022}. Measurements were taken with 30~mJ laser pulses and 8~W of heating laser power, which was observed to optimize the peak atomic density. $5.3(5) \times 10^{10}$ SrOH molecules were produced using the thermal emission method with a peak density of $3.0(3)\times 10^{8}$ molecules/cm$^3$, comparable to yields of ablation-based sources reported in previous experimental results~\cite{kozyryev_radiation_2016,kozyryev_sisyphus_2017}. In more recent work with cryogenic beams of alkaline earth(-like) hydroxide molecules, it has been shown that laser excitation to the $^3P_1$ atomic state can increase molecular yield by an order of magnitude~\cite{jadbabaie_enhanced_2020}. We expect that directing $\sim$0.1--1~W of atomic excitation light into the cell may improve molecular yield in a thermal emission source by an even larger factor, since atoms produced via thermal emission (unlike those produced via ablation) are not expected to significantly populate excited electronic states where chemical reactions are favored. High-power atomic excitation light was not available for this work, but future studies investigating molecular production in the presence of atomic laser excitation are strongly motivated.

\section{Discussion}

This work was inspired by previous reports of loading a MOT of Sr atoms via laser light impinging on SrO~\cite{kock_laser_2016, hsu_laser-induced_2022}. However, the mechanism responsible for this phenomenon has so far not been fully illuminated. Here we present the features of SrO that make it suitable for thermal production of Sr, and summarize several experimental indications of a thermal production mechanism.

Typical of ceramics, SrO possesses a relatively low thermal conductivity of 11~W$/$(m$\cdot$K$)$ around room temperature~\cite{szelagowski_effective_1999}, which enables a tightly focused laser to raise a small volume of SrO to temperatures of likely >$10^{3}$~K before the thermal energy can diffuse to the bulk material. A toy model of the heat dissipation in the mixed SrO and HfC target is presented in App.~\ref{sec:temperature}, resulting in an estimated temperature near the 2800~K melting point of SrO~\cite{irgashov_thermodynamic_1985}, where the vapor pressure becomes sufficiently large to compete with ablation-based production, when impinged on by a 50~$\mu$m diameter laser beam at several watts. However, this would be of little advantage if SrO did not also possess the fortuitous property of dissociating into Sr(g) and O(g) at high temperatures~\cite{asano_partial_1972}. Ideally, a material used for laser-induced thermal production would also have a high absorption efficiency, which we achieve by mixing in a fine powder of the refractory, inert, dark ceramic HfC.

The foregoing properties are required for efficient thermal production, but do not rule out a significant contribution from a non-thermal mechanism such as those responsible for light-induced atomic desorption of alkali atoms~\cite{meucci_light-induced_1994, gozzini_light-induced_1993, rebilas_comment_2010, rebilas_light-induced_2009, barker_light-induced_2018, alexandrov_light-induced_2002, anderson_loading_2001, zhang_light-induced_2009}. Several observations strongly suggest a predominantly thermal mechanism in our apparatus. First, the emission effect is strongly dependent on the laser focus: adjusting the position of the focusing lens by a few mm away from its optimum typically eliminates observable atom flux. Second, the total emission is strongly dependent on laser power, not only the total number of incident photons (i.e., deposited energy). Third, darkening the target with HfC powder dramatically improves the production of strontium: attempted thermal emission of a pressed powder pellet of pure, white SrO does not result in a significant atom flux (i.e., within $\sim$10\% of typical ablation pulses) even at the highest available laser powers.

Ablation-based sources have been implemented in a wide variety of experimental atomic, molecular, and optical (AMO) physics settings~\cite{acme_collaboration_improved_2018, hutzler_buffer_2012}. We suggest that the thermal emission method may achieve higher fluxes in more diverse chemical species of atoms and molecules beyond strontium. We note that this experiment is closely related to prior work producing ThO(g) from a laser-heated mixture of Th and ThO$_{2}$ powders in a CBGB~\cite{petrik_west_thermochemical_2017}. The same method as described here may also be applicable to other alkaline earth oxides to produce alkaline earth(-like) species of great interest including Mg, Ca, Ba, Ra, and Yb~\cite{takahashi_engineering_2023, kozyryev_precision_2017, vilas_magneto-optical_2022, hallas_optical_2023, anderegg_quantum_2023, isaev_laser-coolable_2017, chae_entanglement_2021, norrgard_radiative_2023, altuntas_demonstration_2018, the_nl-eedm_collaboration_measuring_2018}. Furthermore, certain thermally insulating pure metals may prove suitable for laser-induced thermal production. For example, Ba, Ra, Dy, and Y are all elements of interest to ongoing AMO experiments which possess thermal conductivities below 20~W$/($m$\cdot$K$)$, comparable to SrO~\cite{isaev_laser-coolable_2017, arrowsmith-kron_opportunities_2024, leefer_towards_2014, collopy_3d_2018, the_nl-eedm_collaboration_measuring_2018, altuntas_demonstration_2018}.

Additionally, in special cases such as with rare isotopes or radioactive elements like Ra, efficient conversion from precursor material to gas-phase atoms is highly desirable. Ablation introduces inefficiencies to the vaporization process, due to the production of dusts and droplets of ablated material detritus~\cite{chichkov_femtosecond_1996}. An alternative method that directly produces gas-phase atoms, for example by laser heating of RaO or the thermally insulating Ra metal, may therefore offer significant technical advantages.

Possibilities also exist for alternative target preparation methodologies beyond this work. For example, a target prepared from finer powders, with grain sizes much smaller than the focused laser spot size, may improve spot-to-spot production stability. A more speculative approach would be to deposit a thin layer of SrO on a thin substrate of a dark thermal insulator, which could be directly heated from behind. Such an approach might achieve both more efficient laser power absorption by the dark layer and improved local heating of the SrO layer compared to the bulk mixed-material target studied in this work.

\section{Conclusion}

We have demonstrated a high-flux source of cold atoms and molecules that employs laser heating of an SrO target in a cryogenic buffer gas beam source. With 30 mJ laser pulses, a typical value for experiments with cryogenic molecular beams, peak atomic density of the thermal emission source is up to $1.93(6) \times 10^{12} \text{ atoms}/\text{cm}^3$, twice that of a comparable ablation reference, while the total atomic yield is up to $3.7(2) \times 10^{14}$~atoms, 7$\times$ that of the ablation reference. The improved atomic yield of this source compared to an ablation-based source, combined with the observation of molecular production already on par with conventional ablation yields, suggests a direct benefit to molecular beam experiments with strontium-containing molecules in future work where high-power atomic excitation light can better facilitate chemical reactions in the cryogenic cell. The thermal emission technique may also be extended to produce other elements in a cryogenic source by direct heating of metals with relatively low thermal conductivity or other alkaline earth oxides.

\begin{acknowledgments}

\textit{Acknowledgements.} We are grateful to Mingda Li for constructing part of the apparatus used for a preliminary investigation of this topic, as well as Alexander Frenett and Yicheng Bao for technical assistance and useful discussions. This work was done at the Center for Ultracold Atoms (an NSF Physics Frontiers Center) and supported
by Q-SEnSE Quantum Systems through Entangled Science
and Engineering (NSF QLCI Award No. OMA-2016244).

\end{acknowledgments}

\begin{appendix}

\section{Simple model of laser heating}
\label{sec:temperature}

To estimate the temperature range of the surface of the mixed SrO and HfC target, we consider a simple toy model similar to that employed in~\cite{petrik_west_thermochemical_2017} to estimate the temperature of a laser-heated pressed powder target of Th and ThO$_{2}$ in a cryogenic source. We model the laser beam as a uniform disk of radius $R$, measured to be 25~$\mu$m in our system, impinging on a uniform volume of infinite radius and depth. The relevant material properties are the thermal conductivity $k$, density $\rho$, and specific heat capacity $C$, which can be combined into the convenient quantity of thermal diffusivity $\alpha=k/(\rho C)$. Because the laser energy is likely absorbed predominantly in HfC grains, we apply parameters for HfC rather than SrO. Furthermore, although in reality the material properties may vary with temperature, we assume fixed values of $k=27$~W$/$(m$\cdot$K$)$~\cite{opeka_mechanical_1999}, $C=0.7$~J$/$(g$\cdot$K$)$~\cite{savvatimskiy_thermophysical_2020}, and $\rho=12.2$~g$/$cm$^{3}$, giving $\alpha=0.032$~cm$^{2}/$s.

For a laser beam supplying absorbed power $Q$, this model has an analytic solution for temperature over time~\cite{poirier_transport_2016}, which can be simplified at the material surface and the center of the laser beam to give
\begin{equation}
    T(t) = \frac{Q}{\pi R k} \left[\frac{1}{\sqrt{\pi}}\sqrt{\frac{t}{\tau}}\left(1-e^{-\tau/t}\right)+\mathrm{erfc}\left(\sqrt{\frac{\tau}{t}}\right)\right],
\end{equation}
where $\tau=R^{2}/(4\alpha)\approx50~\mu$s is a characteristic timescale for the heating process. The quantity in brackets smoothly increases from 0 to 1 as $t\rightarrow \infty$ so that the temperature asymptotically approaches $Q/(\pi Rk)$. In this model, we find that with 5~W of absorbed laser power, the surface heats to a maximum temperature around 2350~K, reaching half this temperature in $1\tau\approx50~\mu$s and 90\% of the maximum temperature in about 31$\tau\approx1.6$~ms. One may therefore expect SrO grains adjacent to hot HfC grains to conductively heat to temperatures $\gtrsim1000$~K, where the vapor pressure becomes substantial.

This model is intended to provide intuition for the range of temperatures that might reasonably be achieved in our system. The most severe limitations on the model are the effect of a mixed-compound target material, the fact that the laser beam waist is comparable to the powder grain diameter, and the lack of available thermal property data covering the entire relevant range of temperatures. However, it serves to support the thermal emission model by illustrating the plausibility of reaching SrO temperatures around 1000~K or higher in timescales on the order of 0.1$-$1~ms of heating with a tightly focused, high-power laser.

\section{Calculation of emitted atom number}
\label{sec:atom_number}

We describe here how the number density of atoms in the cell over time, $n(t)$, can be used to calculate the total number of produced atoms, $N$. We assume that atoms are produced at some unmeasured rate $r(t)$ so that $N = \int r(t) dt$. Let the atomic number density in the cell following an instantaneously produced atom at $t=0$ be given by $p(t)$, so that $n(t)=r(t) * p(t)$ where $*$ denotes a convolution of functions. Then by the properties of convolutions, $\int n(t) \mathrm{d}t = \int r(t) \mathrm{d}t \times \int p(t) \mathrm{d}t$ and it follows that $N = \int n(t) \mathrm{d}t/\int p(t) \mathrm{d}t$.

Since $p(t)$ is the in-cell atomic number density for a single atom produced at $t=0$, we must have $p(t=0)=1/V$ where $V$ is the volume of the cell. Furthermore, since ablation is a nearly instantaneous atom production event, we can infer the time dependence of $p(t)$ by observing the decay of in-cell density following ablation (see Fig.~\ref{fig:time_traces}). The density is well-described by a single decaying exponential with characteristic time $\tau = 2.85(2)$~ms, which we fit from measurements of atomic density following ablation. Thus $p(t)=e^{-t/\tau}/V$ and we find that $N=(V/\tau)\int n(t)\mathrm{d}t$. Indeed, one can show (though we omit the details here) that this expression for the total produced atomic number is valid for \emph{any} reasonably well-behaved functional form of $p(t)$ provided $\tau$ is interpreted as the mean time for an atom to exit the cell.

\end{appendix}

\bibliographystyle{custom-apsrev4-2}
\bibliography{sr_thermal}

\end{document}